\documentstyle[12pt,epsf]{article}
\begin{document}
\vspace*{1cm}

\vskip 0.5cm
\begin{center}
{\Large \bf Asymmetry of prompt photon
production \\[0.2cm]
 in  $\vec p-\vec p$ collisions at RHIC}
\end{center}

\vskip 1cm
\begin{center}

{\bf G. P. \v {S}koro}\footnote{goran@rudjer.ff.bg.ac.yu},
\vskip 0.3cm
{\it Institute of Nuclear Sciences "Vin\v {c}a", \\
Faculty of Physics, University of Belgrade,\\
Belgrade, Yugoslavia}

\vskip 0.5cm

{\bf M. Zupan}\footnote{mzupan@rt270.vin.bg.ac.yu},
\vskip 0.3cm
{\it Institute of Nuclear Sciences "Vin\v {c}a", \\
Belgrade, Yugoslavia}

\vskip 0.5cm

{\bf
 M. V. Tokarev}\footnote{tokarev@sunhe.jinr.ru}
\vskip 0.3cm
{\it Laboratory of High Energies,\\
Joint Institute for Nuclear Research,\\
141980, Dubna, Moscow region, Russia}
\end{center}

\vskip 2cm

{\bf{Summary.$-$}\it{
The prompt photon production in $\vec p-\vec p$ collisions at high energies
is studied. Double-spin asymmetry $A_{LL}$ of the process is calculated
by using Monte Carlo code SPHINX. A possibility
to discriminate the spin-dependent gluon  distributions
and to determine sign of $\Delta G$ is discussed.
Detailed study of expected background, such as $\pi^{0}$ production
and decay, is given. The predictions for the longitudinal
asymmetry $A_{LL}$ of the prompt photons and $\pi^{0}$-meson
 production in the
$\vec p-\vec p$  collisions at RHIC  energies have been made.
}} \\[5mm]

\newpage

{\large {\bf 1 $-$ Introduction}}
\vskip 0.3cm

One of the actual problems of high energy spin physics is the measurement
of the spin-dependent gluon distribution $\Delta G(x,Q^2)$. This would
allow the determination of gluon and moreover
of quark contribution to the spin of the proton. Although spin-dependent
quark distributions were
obtained from deep-inelastic lepton-nucleon scattering, this information
was not sufficient to
determine the various parton contributions to the proton spin
(see \cite{Leader} and references therein).

Consequently the preparation of many experiments is well under way.
These experiments are
to take place on colliders that are presently under construction,
such as RHIC \cite{RHIC},
HERA \cite{HERA} and LHC (see \cite{LHC} and references therein).
These colliders will provide high-energy polarized proton beams necessary
for the study of the
asymmetry of jet and/or prompt photon production which can be used as
a measure of  $\Delta G(x,Q^2)$.
In a recent paper \cite{jet}, we analyzed asymmetry of jet and dijets
production at RHIC by means of Monte Carlo simulations and found that
the asymmetry $A_{LL}^{jet}$ is sensitive  for $\Delta G$ and can give
us the information about the sign and shape of spin-dependent
gluon distribution.
In this paper we study prompt photon production and the asymmetry
of the production with
respect to the parallel and anti-parallel orientation of the
longitudinally oriented spins of the colliding protons.

The processes responsible for prompt photon production are the
Compton $qg {\rightarrow}q{\gamma}$ scattering and the
annihilation process  $q\bar {q} {\rightarrow}g{\gamma}$.
The asymmetry of prompt photon production will therefore
be dependent on the
convolution of
spin-dependent distributions of quarks and gluons. The set of possible
distributions is subject to
very few constraints. Additional ones must be arrived at theoretically
or empirically, or both. For
example, the sign of the gluon spin contribution is subject to speculation.
Although there are
indications that it should be positive, the possibility that it may be
negative also exists.
Both positive and negative values of
the sign  of  $\Delta G(x, Q^2)$
were considered in \cite{tok1}.
The possibility to draw conclusions on the sign of the
spin-dependent gluon distribution, $\Delta G(x, Q^2)$, from existing
polarized DIS data have been studied in \cite{tok2}.
Other speculations deal with the relative magnitude of the gluon
contribution, models are
proposed in which $\Delta G$ is large compared to other contributions,
others where it is small
compared to other contributions.

All this points to the necessity of  measurement of $\Delta G(x, Q^2)$,
and consequently $\Delta G$. The goal of this paper
is to study the asymmetry of
prompt photon production in proton-proton collisions
for positive and negative $\Delta G(x, Q^2)$ and to predict the magnitude
of the effect when observed by
the STAR detector at RHIC \cite{Bunce}. Moreover, the paper is intent on
presenting some of the
considerations involved in the selection of the kinematical region which
is optimal for the
observation of the effect. Finally some consideration was made of the
expected background
giving a prediction of the asymmetry signal that would actually be observed
in the experiment.

\vskip 1cm

{\large {\bf 2 $-$ Spin-dependent gluon distribution}}
\vskip 0.3cm

A phenomenological spin-dependent distributions \cite{tok1} were used in
the calculation of the
prompt photon production asymmetry.  The distributions \cite{tok1} were
arrived at empirically,
including some constraints on the signs of valence and
sea quark distributions, taking  into account the axial
gluon anomaly and utilizing results on integral quark contributions
to the nucleon spin.  Based on the analysis of experimental
deep inelastic scattering data for the structure function $g_1$
the parametrizations of
spin-dependent parton distributions for both
positive  and negative sign of  $\Delta G$ have been constructed.
We would like to note that both sets of distributions describe
experimental data very well.
We shall denote  $\Delta G^{>0}$  and   $\Delta G^{<0}$
sets of spin-dependent parton distributions obtained
in \cite{tok1}  with positive and negative
sign of $\Delta G$, respectively.
It was shown in \cite{tok3} that
the constructed spin-dependent parton distributions
for positive sign of $\Delta G$
reproduce the main features of the NLO QCD $Q^2$-evolution
of proton, deuteron  and neutron  structure function $g_1$.

Figure 1 shows the dependence  of the ratio $\Delta G(x,Q^2)/G(x,Q^2)$
on $x$ at $Q^2=100\ (GeV)^2$ for gluon distributions
$\Delta G^{>0}$ and $\Delta G^{<0}$ \cite{tok1}. In the case of
positive $\Delta G(x,Q^2)$
 we see a monotonically increasing curve corresponding to the behaviour
$\Delta G/G\sim x^{\alpha}$  as $x \rightarrow 1$. For $\Delta G^{<0}$
we see a monotonically
decreasing curve behaving almost exactly as $\Delta G/G\sim -x^{1/2}$, as
$x \rightarrow 1$.
Figure 2 shows the dependence  of the ratio $\Delta q(x,Q^2)/q^{R}(x,Q^2)$
on $x$ at $Q^2=100\ (GeV)^2$ for u-quark (a) and d-quark (b).
We would like to note that $q^{R}(x,Q^2)$ represents the renormalised
parton distribution
as given in \cite{tok1}.


\vskip 1cm

{\large {\bf 3 $-$ Asymmetry of prompt photon production}}
\vskip 0.3cm

There are two principal processes that produce prompt photons: the
Compton process $qg {\rightarrow}q\gamma$ scattering and the annihilation process  $q\bar {q} {\rightarrow}g\gamma$.
However for large gluon polarisations the contribution of
the annihilation process can be neglected.

The longitudinal, double spin asymmetry is defined as the
difference of cross-sections for
prompt photon production when the longitudinally oriented spins of
the colliding protons are
antiparallel (${\uparrow}{\downarrow}$) and parallel
(${\uparrow}{\uparrow}$):

\begin{equation}
A_{LL}=\frac{\sigma^{{\uparrow}{\downarrow}} -
\sigma^{{\uparrow}{\uparrow}}}{\sigma^{{\uparrow}{\downarrow}}
+ \sigma^{{\uparrow}{\uparrow}}}
 ={\frac{1}{P^2}}{\cdot}
 \frac{N_{\gamma}^{{\uparrow}{\downarrow}} -
N_{\gamma }^{{\uparrow}{\uparrow}}}{N_{\gamma}^{{\uparrow}{\downarrow}}
+ N_{\gamma}^{{\uparrow}{\uparrow}}},
\end{equation}

where $N_{\gamma}$ represents the number of prompt photons.
The statistical error is calculated as:

\begin{equation}
{\delta}A_{LL}{\simeq}{\frac{1}{P^2}}{\cdot}\frac{1}{\sqrt{N_{\gamma}^{{\uparrow}{\downarrow}}
+ N_{\gamma}^{{\uparrow}{\uparrow}}}}.
\end{equation}

The calculation of prompt photon production asymmetry was done using
Monte Carlo code SPHINX \cite{SPHINX} which is a 'polarized' version of
PYTHIA \cite{PYTHIA}. Calculations of asymmetry were made for center-of-mass
energies $\sqrt s=200~GeV$
and $\sqrt s=500~GeV$ which are part of the design specifications of RHIC.
The STAR detector is designed to cover full space
in azimuth and pseudorapity region $-1<{\eta}<2$, so this was
taken into account in calculating the asymmetry of prompt photon production.

The rate and asymmetry of prompt photon production was estimated using the
assumed RHIC integrated
luminosity $320~pb^{-1}$ at $\sqrt s = 200\ GeV$  and $800~pb^{-1}$ at
$\sqrt s = 500\ GeV$
and a fixed beam polarization of  P = 0.7.

\vskip 1cm

{\large {\bf 4 $-$ Results and discussions}}
\vskip 0.3cm

The asymmetry of prompt photon production was calculated from simulations
for both
$\Delta G^{>0}$ and $\Delta G^{<0}$ sets of spin dependent PDF \cite{tok1}.
 Figure 3 shows the dependence of $A_{LL}$ on the photon transverse
 momentum
$p_T$ at $\sqrt s=200~GeV$. For $\Delta G^{>0}$ the asymmetry increases
from about 1.5\% at $p_{T} = 6 ~GeV/c$ to 13\% at $p_{T} = 24~ GeV/c$.
For $\Delta G^{<0}$ the asymmetry drops from zero
at $p_{T} = 6 ~GeV/c$ to -17\% at $p_{T} = 24~ GeV/c$. The errors shown
are statistical.
Figure 4 shows the asymmetry of prompt photon as a function of prompt
photons transverse momentum
$p_T$ at $\sqrt {s} = 500~GeV$. This asymmetry is much
smaller than the one at $\sqrt{s} = 200~GeV$. It is practically equal
to zero for $p_T$ up to about $14 GeV/c$.
Beyond  $p_{T} = 14~ GeV/c$ it rises slowly for $\Delta G^{>0}$ and drops
for $\Delta G^{<0}$.

This means that the energy of  $\sqrt s = 200\ GeV$ will be more
preferable for determination of
$\Delta G$ from prompt photon production. Note that higher colliding
energy $\sqrt s = 500\ GeV$
is preferable for extracting $\Delta G$ from jets asymmetry \cite{jet}.
Figure 5 shows estimated rates of prompt photon production with above
mentioned conditions
at $\sqrt s = 200\ GeV$ and $\sqrt s = 500\ GeV$.

The rate of prompt photon production decreases with
increasing $p_T$ due largely to the decrease in the distribution of
gluons at high $x$, keeping in mind
the approximate relation $x = p_T/(2\sqrt {s})$. At the same time the
asymmetry effect increases with $p_T$.
Consequently the successful measurement of the asymmetry  in
prompt photon production
depends largely on reconciling the magnitude of the effect with the
statistical reliability of the
measurement. For example, at $\sqrt s = 200~GeV$ only the asymmetry
values up to
$p_{T} {\simeq} 25~ GeV/c$ will have acceptable statistical significance.
The problem of experimental errors
have to be analyzed in the light of corresponding background processes, too.

The main source of background, in this case, is the production of $\pi^{0}$
mesons because the $\pi^{0}$'s
general mode of decay $\pi^{0}{\rightarrow}2\gamma$
significantly affects prompt photon detection.
High-energy $\pi^{0}$'s decay into photons diverging mostly at
an angle $\theta$,
given by $sin ({\theta}/2)=m_{\pi^{0}}/ E_{\pi^{0}}$, small enough to be
collected by a single cell of the Electro-Magnetic Calorimeter and
counted as a high-$p_T$ prompt photon. Also the possibility of one
of the decay photons escaping detection causes the one which is
detected to be considered a prompt photon.

Also, the processes responsible for $\pi^{0}$-meson production are
$q-q$, $q-g$ and $g-g$
scattering, so in the polarized $pp$ collisions non-zero values of
$\pi^{0}$ asymmetry can be expected. Figure 6 shows the asymmetry of
$\pi^{0}$-meson production as a function of transverse momentum $p_T$ at
$\sqrt s = 200\ GeV$. The asymmetry is positive and increase with
$p_T$ for $\Delta G^{>0}$ and practically equal to zero for
$\Delta G^{<0}$. Such a behaviour will reflect on the expected experimental
prompt photon asymmetry.
Also, we can see in Figure 7 that the average cross-section of
$\pi^{0}$-meson
production is higher than
average cross-section of prompt photon production by an order of magnitude
in the whole kinematical
region. To provide an adequate prediction of the asymmetry measured
in the experiment, the
production of $\pi^{0}$'s has to be taken into account in the
calculation of $A_{LL}$.  Also, in order to reduce such a high background,
the optimal method for the so-called "$\pi^{0}/{\gamma}$ separation"
should be performed. The very important part of the EM Calorimeter at
STAR is Shower Maximum Detector which can be used for the purpose \cite{SMD}.
The method described in Ref. \cite{SMD} is based on the different shapes
of the EM showers induced by gammas and pions.
The experimental asymmetry was calculated by substituting:

\begin{equation}
N_{\exp }^{{\uparrow}{\uparrow}}={\it a}{\cdot}{N_{\gamma}^{{\uparrow}{\uparrow}}
+ {\it b}{\cdot}N_{\pi^{0}}^{{\uparrow}{\uparrow}}},
\end{equation}
for $N_{\gamma}^{{\uparrow}{\uparrow}}$ into (1), and by an analogous
substitution for $N_{\gamma}^{{\uparrow}{\downarrow}}$.
The quantity $N_{\pi^{0}}^{{\uparrow}{\uparrow}}$
is the number of produced pions
and $\it a$ and $\it b$ are constants determined by
the gamma detection efficiency and the percent of misidentified pions
, respectively.
These constants were taken to be: ${\it a} = 0.9$ reflecting a
90\% photon detection efficiency and ${\it b} = 0.35$ reflecting
a 65\% efficiency in $\pi^{0}$-meson rejection from
the signal as shown in \cite{SMD}.

Figure 8 shows the 'experimental' prompt photon asymmetry obtained
by taking into account all conditions described above.
The experimental asymmetry is different for different signs of
$\Delta G$ implying that it will be a clear signal at least for
the sign of $\Delta G$.

\vskip 1cm

{\large {\bf 5 $-$ Conclusions}}

\vskip 0.3cm
Monte Carlo simulations of prompt photon production in polarized
proton collisions at high
energies were made, taking into account parameters of the STAR detector
at RHIC.  The
dependence of asymmetry of prompt photon production on the transverse
momentum of photons
was studied for positive and negative values of $\Delta G$ at colliding
energies of   $\sqrt s = 200\ GeV$
and  $\sqrt s = 500\ GeV$. The results show that
a stronger asymmetry signal with a
clear indication of the sign of $\Delta G$ will be at
$\sqrt s = 200\ GeV$. At that energy the
asymmetry $A_{LL}$ ranges from  1.5\% to 13\% for the $p_T$ range of
$6~GeV/c$ to $24~GeV/c$.
The asymmetry effect is larger at higher $p_T$ values but at the
same time the number of prompt photons produced at high $p_T$ is smaller.
The measurement of the asymmetry will need to
reconcile the visibility of the effect and the statistical reliability of
the measurement.
The decay of $\pi^{0}$ mesons was considered as a source of background.
It was found that the
number of photons from $\pi^{0}$ decay can exceed the  number of prompt
photons by an order of magnitude. Taking into account  the results on
$\pi^{0}/{\gamma}$ separation achievable at STAR,
 it was shown that the experimental asymmetry will also give a clear
 signal as to the sign of $\Delta G$.

\vskip 0.5cm



{


}

\vskip 0.5cm
\newpage
\begin{minipage}{4cm}
\end{minipage}
\vskip 2cm
\begin{center}
\hspace*{0.5cm}
\parbox{6.5cm}{\epsfxsize=6.5cm \epsfysize=6.5cm \epsfbox[5 5 500 500]
{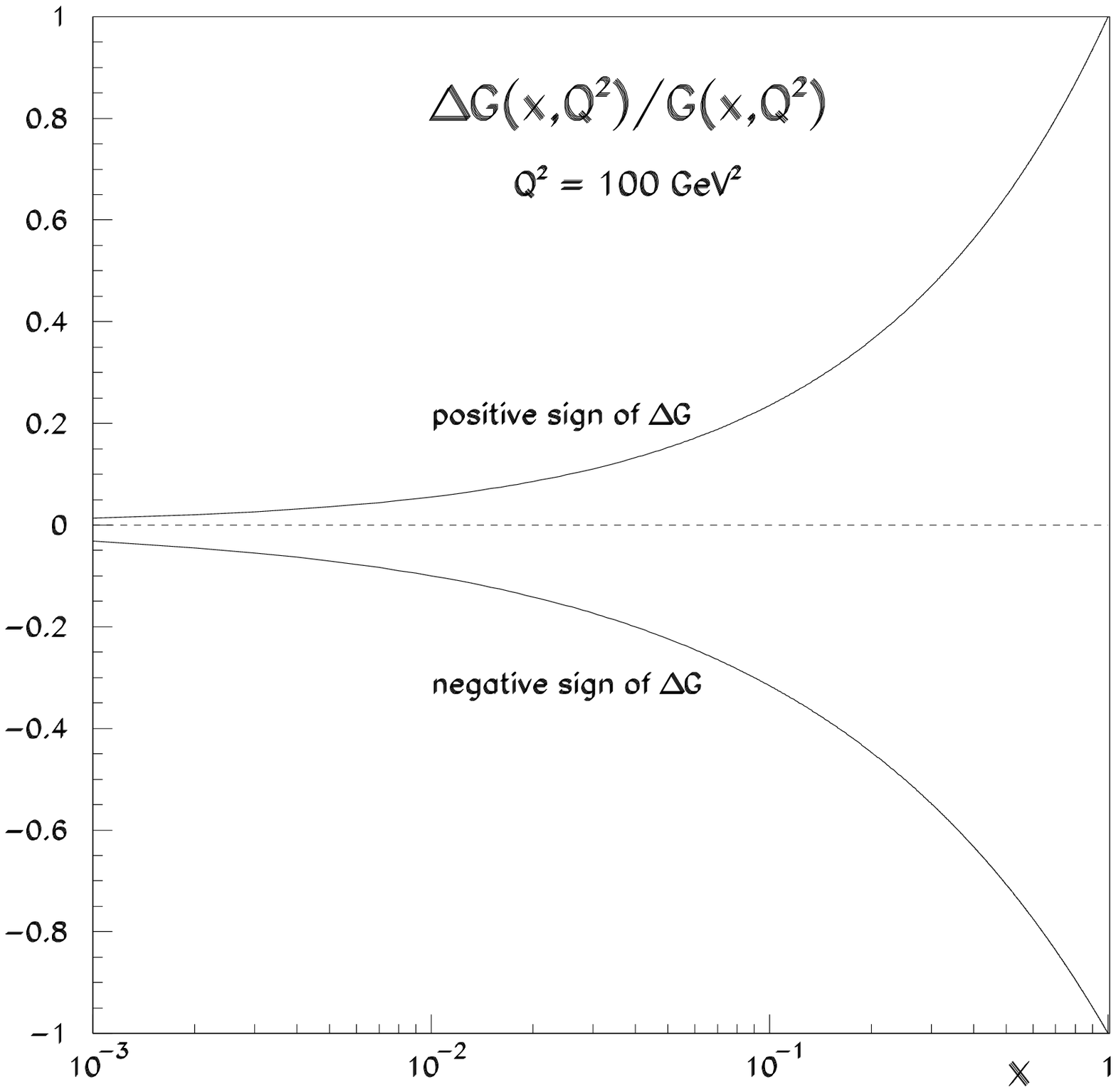}{}}
\end{center}
\vskip -0.5cm

{\bf Figure 1}

The ratio of  polarized and unpolarized
gluon distributions as a function of $x$
for two different parametrizations
$\Delta G^{>0}$ \cite{tok1}  and
$\Delta G^{<0}$ \cite{tok1}    at $Q^2$=100 GeV$^2$.



\vskip 4cm
\begin{center}
\hspace*{-0.5cm}
\parbox{6cm}{\epsfxsize=6.cm \epsfysize=6.cm \epsfbox[5 5 500 500]
{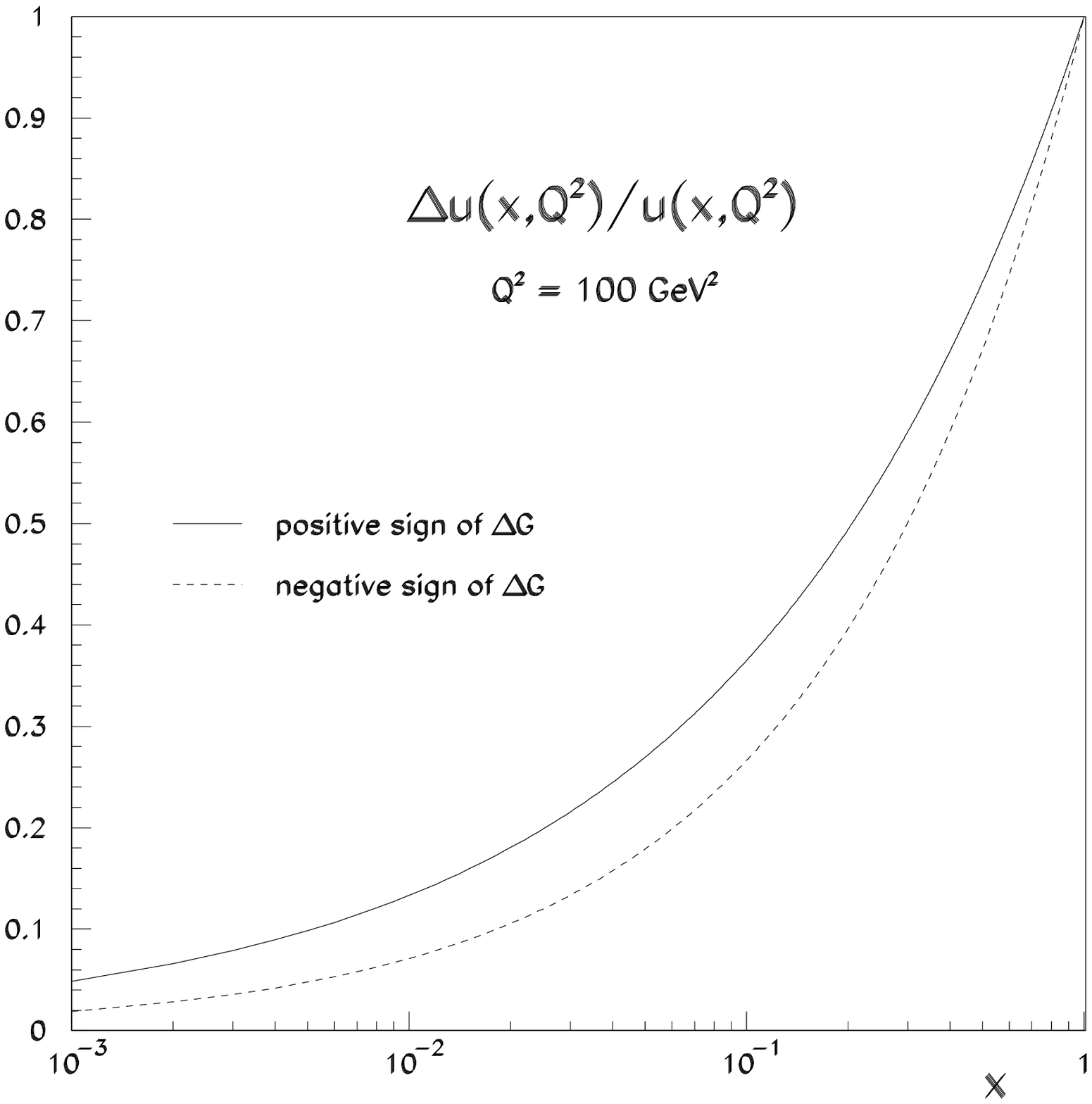}{}}
\hspace*{1cm}
\parbox{6cm}{\epsfxsize=6.cm \epsfysize=6.cm \epsfbox[35 5 535 500]
{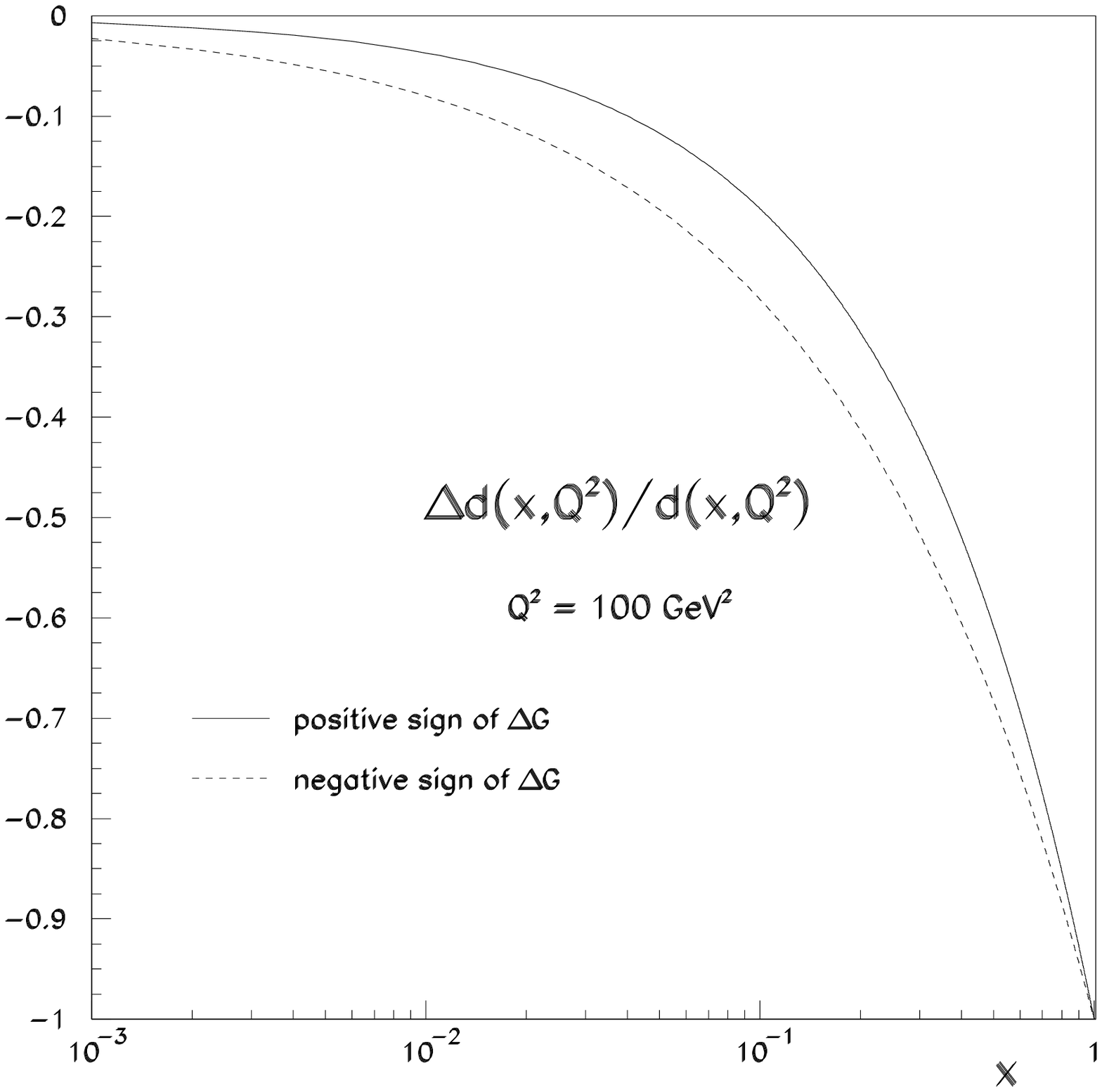}{}}
\vskip -1cm
\hspace*{1. cm} a) \hspace*{5cm} b)\\[0.5cm]

\end{center}
{\bf Figure 2}

The ratio of  polarized and unpolarized
$u$-quark (a) and $d$-quark (b) distributions as a function of $x$
for two different parametrizations
$\Delta G^{>0}$ \cite{tok1}  and
$\Delta G^{<0}$ \cite{tok1}    at $Q^2$=100 GeV$^2$.



\newpage

\begin{minipage}{4cm}
\vskip 2cm
\end{minipage}
\vskip 2cm
\begin{center}
\parbox{6.5cm}{\epsfxsize=6.5cm \epsfysize=6.5cm \epsfbox[35 5 535 500]
{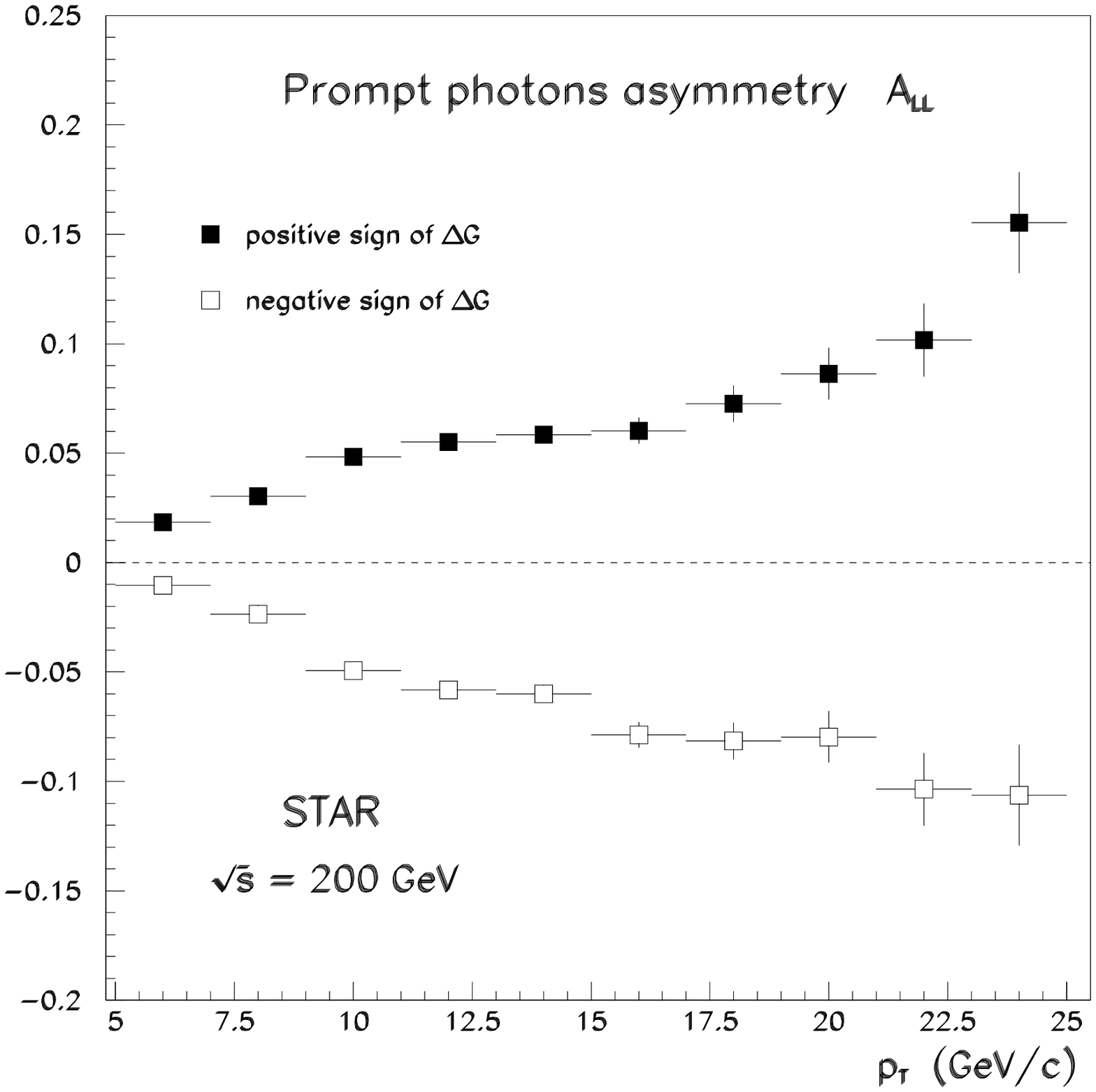}{}}
\end{center}

\vskip -1.5cm

{\bf Figure 3.}

Asymmetry of prompt photon production $A_{LL}$ in polarised $pp$ collisions
at $\sqrt s= 200$~GeV for two different sets of spin-dependent PDFs
(
$\Delta G^{>0}$ \cite{tok1}
and $\Delta G^{<0}$ \cite{tok1})  as a function of photon transverse
momentum $p_T$. The errors indicated are statistical only.

\vskip 2.5cm

\begin{center}
\vskip 1.5cm
\parbox{6.5cm}{\epsfxsize=6.5cm \epsfysize=6.5cm \epsfbox[5 5 500 500]
{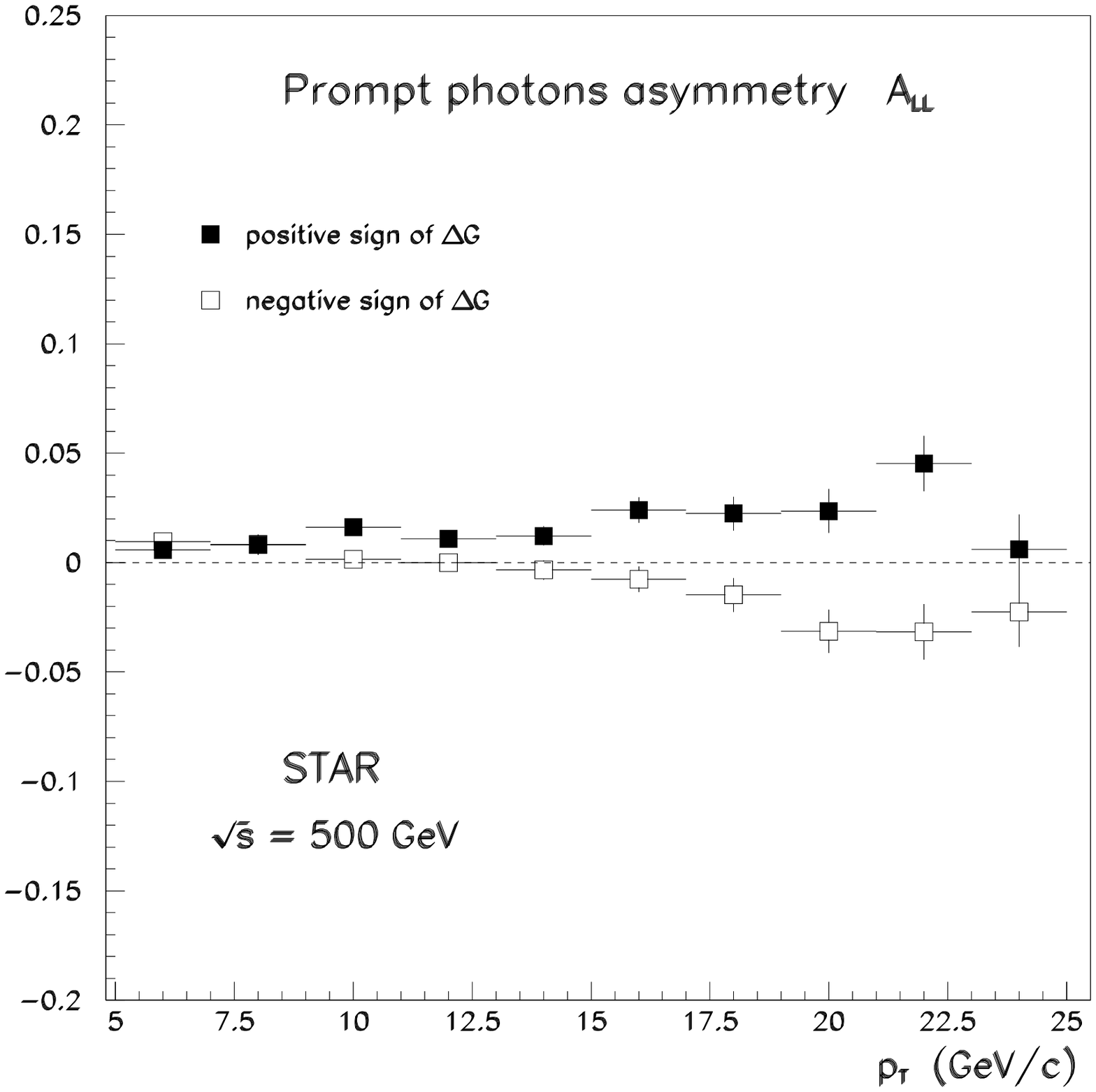}{}}
\vskip -1.5cm
\end{center}
{\bf Figure 4.}

Asymmetry of prompt photon production $A_{LL}$ in polarised $pp$ collisions
at $\sqrt s= 500$~GeV for two different sets of spin-dependent PDFs
(
$\Delta G^{>0}$ \cite{tok1}
 and $\Delta G^{<0}$ \cite{tok1}) as a
function of photon transverse
momentum $p_T$. The errors indicated are
statistical only.


\newpage

\begin{minipage}{4cm}
\vskip  2cm
\end{minipage}
\vskip 2cm
\begin{center}
\parbox{6.5cm}{\epsfxsize=6.5cm \epsfysize=6.5cm \epsfbox[35 5 535 500]
{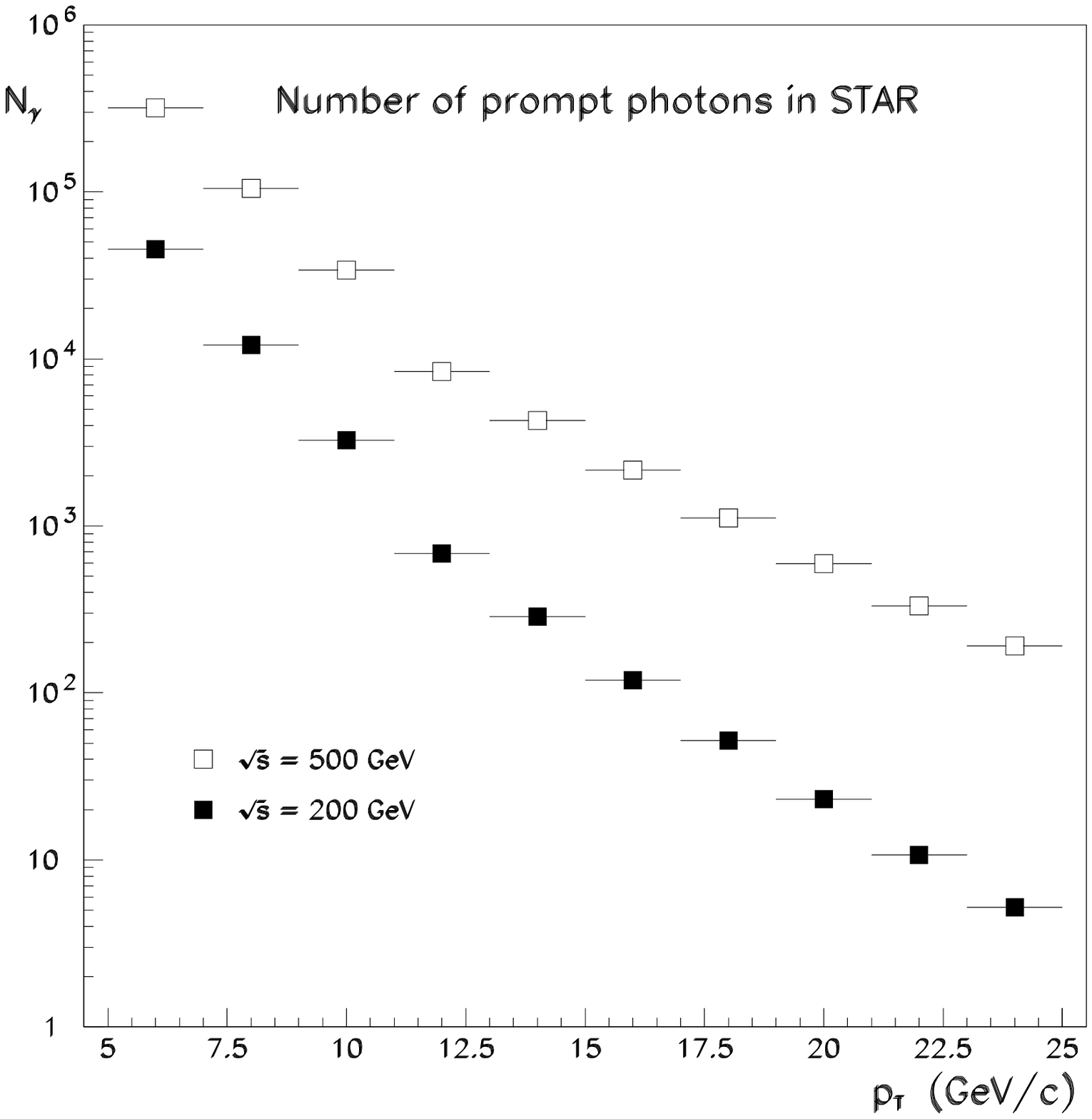}{}}
\end{center}

\vskip -1.5cm

{\bf Figure 5.}

Estimated rates of prompt photon production $A_{LL}$ in polarised $pp$ collisions at $\sqrt s= 200$~GeV and $\sqrt s= 500$~GeV  as a
function of photon transverse
momentum $p_T$. The rates are
based on the expected luminosity of RHIC and the properties of
the STAR detector.

\vskip 2.5cm

\begin{center}
\vskip 1.5cm
\parbox{6.5cm}{\epsfxsize=6.5cm \epsfysize=6.5cm \epsfbox[5 5 500 500]
{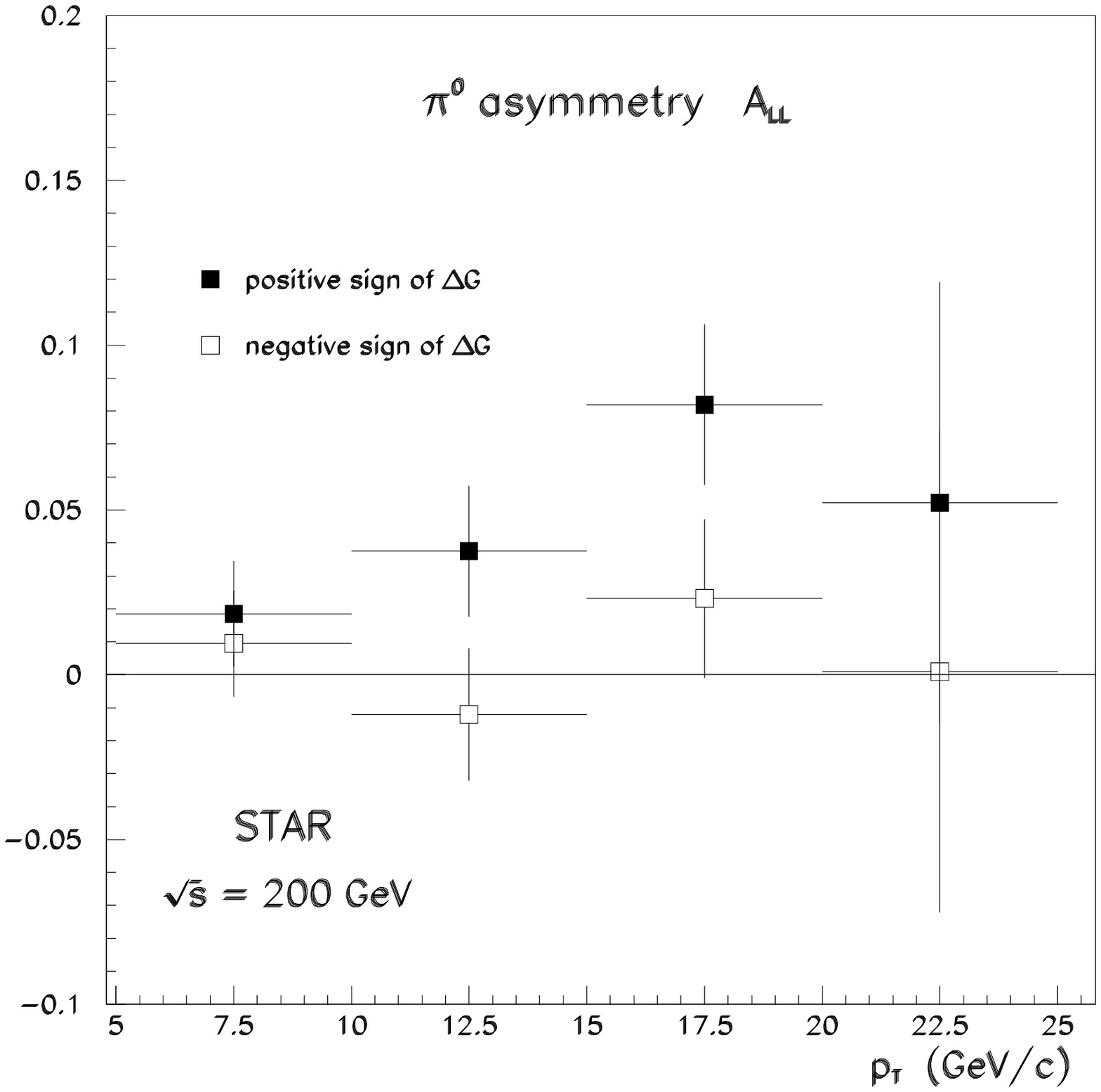}{}}
\vskip -1.5cm
\end{center}
{\bf Figure 6.}

Asymmetry of $\pi^{0}$ production $A_{LL}$ in polarised $pp$ collisions
at $\sqrt s= 200$~GeV for two different sets of spin-dependent PDFs
(
$\Delta G^{>0}$ \cite{tok1}  and
$\Delta G^{<0}$ \cite{tok1}
)
as a
function of $\pi^{0}$ transverse momentum $p_T$. The errors indicated are
statistical only.


\newpage

\begin{minipage}{4cm}
\vskip  2cm
\end{minipage}
\vskip 2cm
\begin{center}
\parbox{6.5cm}{\epsfxsize=6.5cm \epsfysize=6.5cm \epsfbox[35 5 535 500]
{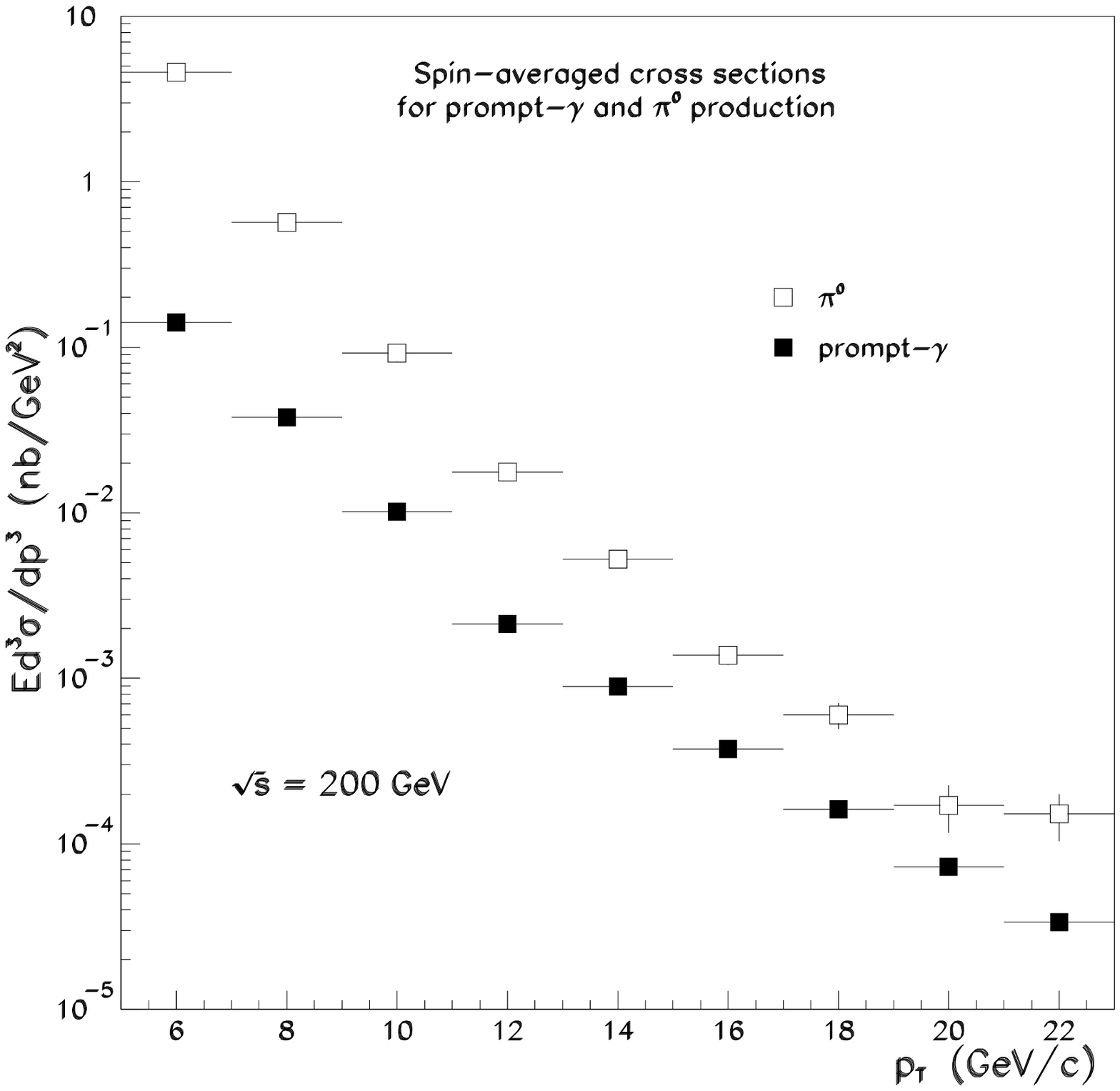}{}}
\end{center}

\vskip -1.5cm

{\bf Figure 7.}

Comparison of prompt photon and $\pi^{0}$ cross-sections  in polarised $pp$ collisions
at $\sqrt s= 200$~GeV  as a function of transverse
momentum $p_T$.
\vskip 2.5cm

\begin{center}
\vskip 1.5cm
\parbox{6.5cm}{\epsfxsize=6.5cm \epsfysize=6.5cm \epsfbox[5 5 500 500]
{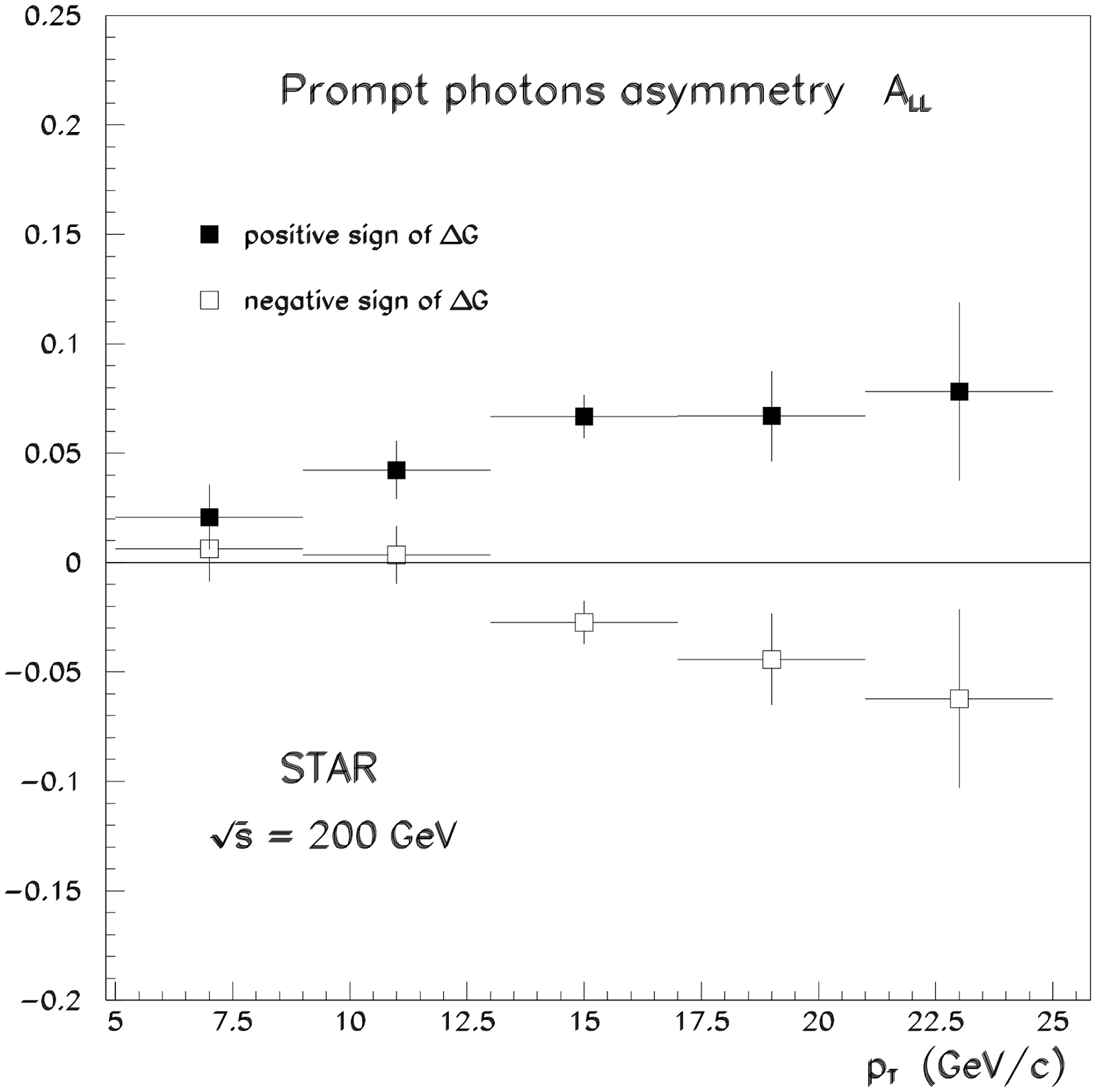}{}}
\vskip -1.5cm
\end{center}
{\bf Figure 8.}

Experimental asymmetry of prompt photon production $A_{LL}$ in polarised $pp$ collisions
at $\sqrt s= 200$~GeV for two different sets of spin-dependent PDFs
(
$\Delta G^{>0}$ \cite{tok1}
and $\Delta G^{<0}$ \cite{tok1})  as a function of photon transverse
momentum $p_T$.
\end{document}